\documentclass[rsi,aip,reprint,superscriptaddress,nofootinbib]{revtex4-1}

\usepackage{amssymb}
\usepackage{amsmath}
\usepackage[ngerman, english]{babel}
\usepackage{graphicx}
\usepackage{natbib}
\usepackage{units}
\usepackage{hyperref}
\usepackage{epstopdf}

\begin{document}

\title{First application studies at the laser-driven LIGHT beamline: Improving proton beam homogeneity and imaging of a solid target} %Title of paper

\author{D. Jahn}
\email[]{d.jahn@gsi.de}
\affiliation{Institut f\"ur Kernphysik, Technische Universit\"at Darmstadt, Schlossgartenstra\ss e 9, 64289 Darmstadt, Germany}
\author{D. Schumacher} 
\author{C.~Brabetz} 
\affiliation{GSI Helmholtzzentrum f\"ur Schwerionenforschung, Planckstra\ss e 1, D-64291 Darmstadt, Germany}
\author{J. Ding}
\author{S. Weih}
\affiliation{Institus f\"ur Kernphysik, Technische Universit\"at Darmstadt, Schlossgartenstra\ss e 9, 64289 Darmstadt, Germany}
\author{F.~Kroll}
\author{F.E.~Brack}
\author{U.~Schramm}
 \affiliation{Helmholtz-Zentrum Dresden - Rossendorf, Bautzner Landstr. 400, 01328 Dresden, Germany} \affiliation{Technische Universit\"at Dresden, 01062 Dresden, Germany}
\author{A.~Bla\ifmmode \check{z}\else \v{z}\fi{}evi\ifmmode \acute{c}\else \'{c}\fi{}} 
\affiliation{GSI Helmholtzzentrum f\"ur Schwerionenforschung, Planckstra\ss e 1, D-64291 Darmstadt, Germany} \affiliation{Helmholtz Institut Jena, Helmholtzweg 4, D-07734 Jena, Germany}
\author{M.~Roth} 
\affiliation{Institus f\"ur Kernphysik, Technische Universit\"at Darmstadt, Schlossgartenstra\ss e 9, 64289 Darmstadt, Germany}

\date{\today}

\begin{abstract}
In the last two decades, the generation of intense ion beams based on laser-driven sources has become an extensively investigated field. The LIGHT collaboration combines a laser-driven intense ion source with conventional accelerator technology based on the expertise of laser, plasma and accelerator physicists. Our collaboration has installed a laser-driven multi-MeV ion beamline at the GSI Helmholtzzentrum f\"ur Schwerionenforschung delivering intense proton bunches in the subnanosecond regime. We investigate possible applications for this beamline, especially in this report we focus on the imaging capabilities. We report on our proton beam homogenization and on first imaging results of a solid target.

\end{abstract}

\pacs{}% insert suggested PACS numbers in braces on next line

\maketitle %\maketitle must follow title, authors, abstract and \pacs
\section{Introduction}
Laser-driven ion acceleration became an extensively investigated field in the past two decades. This promising field, as source of intense ion bunches, is discussed to be used for diverse applications: precise energy-loss measurements in laser-plasmas \cite{witold}, creation of warm dense matter \cite{patelwdm,wdmpelka}, medical applications \cite{nemoto,medicine1,medicine2} or as a diagnostic tool \cite{mackinnon}. In particular, laser-driven proton sources were discussed as a particle probe due to their high brightness and high particle numbers enabling precise time-resolved proton imaging of fast transient phenomena, e.g. imaging of strong electromagnetic fields\cite{mackinnon}, ultraintense interactions and laser-produced plasmas using the continuous energy spectrum\cite{mackinnon2,borghesi}. \\    
The Target Normal Sheath Acceleration (TNSA) mechanism \cite{snavely} is widely used and investigated in the context of laser-driven ion acceleration as it offers excellent beam properties (high intensities, low emittance). Unfortunately, the beam suffers from a large divergence and a continuous broad exponentially decaying energy spectrum, where in contrast most applications demand a monoenergetic collimated beam. Therefore, the Laser Ion Generation, Handling and Transport (LIGHT) collaboration\cite{busold2014} was founded by the universities Technische Universit\"at Darmstadt, Johann-Wolfgang-von-Goethe-Universit\"at Frankfurt, Technische Universit\"at Dresden and the Helmholtz institutes GSI Helmholtzzentrum f\"ur Schwerionenforschung, Helmholtzinstitut Jena, Helmholtz-Zentrum Dresden-Rossendorf. In the past years, the collaboration partners have installed a multi-MeV laser-driven ion beamline, combining a laser-driven ion source with conventional accelerator technology \cite{busold}. \\
\begin{figure}
\includegraphics[scale=0.6]{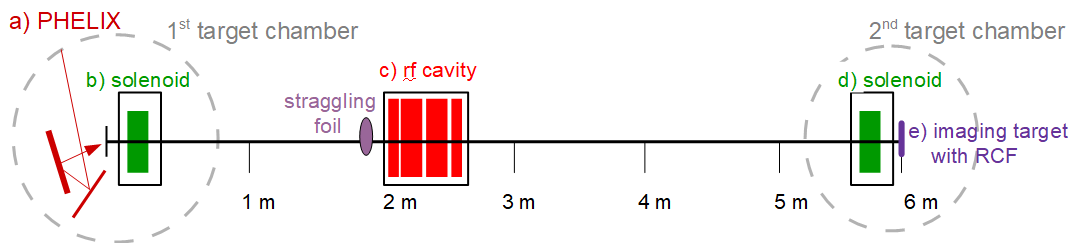}
\caption{The current setup of the LIGHT beamline: a) The PHELIX laser drives the TNSA source delivering a continuous energy spectrum up to 28.4~MeV, b) the high-field solenoid selects and collimates a special energy, c) the bunch is rotated in longitudinal phase space within the rf cavity, d) the bunch is transported to the second target chamber and focused with the second high-field solenoid, and e) the bunch is diagnosed or used for proton imaging.}
\label{beamline}
\end{figure}
The setup of the current LIGHT beamline is shown in figure \ref{beamline}. A sub-aperture beam of the Petawatt High-Energy Laser for Heavy Ion EXperiments (PHELIX) with 100~TW and 5x10$^{19}$~W/cm$^2$ at GSI\cite{Bagnoud2010} drives the TNSA source resulting in a continuous proton energy spectrum up to several tens of MeV. As proton bunches are observed after the use of gold targets for the TNSA mechanism, which are not containing hydrogen, the protons originate from hydrocarbon impurities on the target surface\cite{gitomer1986}.  As the second step, a high-field solenoid captures the generated protons via chromatic focusing, in our case in the range 9$\pm$1 MeV\cite{busold2013}. Additionally, a 1.25 $\mu$m mylar straggling foil is used for homogeneity improvement of the filamented beam after the chromatic focusing stage. The collimated beam is injected into a synchronous radiofrequency (rf) cavity (operated at 108.4~MHz and up to an electric potential of $\pm$~1~MV) enabling longitudinal phase rotation. Then it is transported to a second target chamber located 6~m behind the source. In this chamber, diagnostics and the imaging target are positioned. This beamline design yields proton bunches with more than 10$^{8}$ particles per bunch which can be energy-compressed or time-compressed into the subnanosecond regime, reported by S. Busold \textit{et al.} \cite{busold}. \\
These particular properties of the generated LIGHT proton beams (high brightness, energy-compressed or time-compressed bunch) make them interesting for a variety of possible applications, especially imaging techniques. This report focuses on improving the beam homogeneity and using a solid target for first imaging measurements.

\section{Characterization of the proton beam homogeneity}
Going towards imaging applications, a uniform beam profile is necessary. The collimated LIGHT proton beam covers a circular area with a 40 mm diameter due to the solenoid's  aperture. Therefore, the beam homogeneity using different beamline setups was investigated. The proton detector used in these measurements consisted of several spatially resolving radiochromic (RCF) dosimetry films (\textit{GafChromic}, type:~EBT3) and guarantees spectral and high spatial resolution.  
These dosimetry films measure the radiation dose by colour changes from transparent to blue within the film through the interacting ionizing process. These films are scanned with a Nikon Super Coolscan 9000 ED  and converted the blueing into optical density \cite{nuernberg}. In the next step, the optical density is converted into deposited energy of protons, based on calibration data. Arranging the RCFs in a stacked configuration allows  to determine the spatially resolved energy distribution.  \\% stopping range of protons? Target magnitude smaller?
For the homogeneity analysis, we adapted two parameters which describe the homogeneity of our particle beam: \\
The fill factor (FF) gives a measure for the uniformity of a beam. It is defined as the ratio of the beam volume with an intensity above a threshold level and the volume of an enclosing cylinder calculated with the maximum value. For a perfect top-hat the fill factor has a value of 1. The fill factor is sensitive to small areas with high intensities (\glqq hot spots\grqq). In this work, the threshold was chosen as one percent of the arithmetic average.   \\
The beam uniformity $U_{\eta}$ is defined as the normalized root mean square deviation of deposited energy density from its average value\cite{iso}: % (DIN ISO 13694:2015):
\begin{equation}
U_{\eta} = \frac{1}{E_{\eta}} \sqrt{\frac{1}{A_{\eta}}\int \int [E(x,y) - E_{\eta}]^2 dx dy}
\end{equation}
with the local deposited energy density $E(x,y)$, the deposited average energy density $E_{\eta}$ and the effective irridiation area $A_{\eta}$. In case of $U_{\eta}$ = 0, the beam is completely uniform. \\
\begin{figure}[htp!]
\includegraphics[scale=0.5]{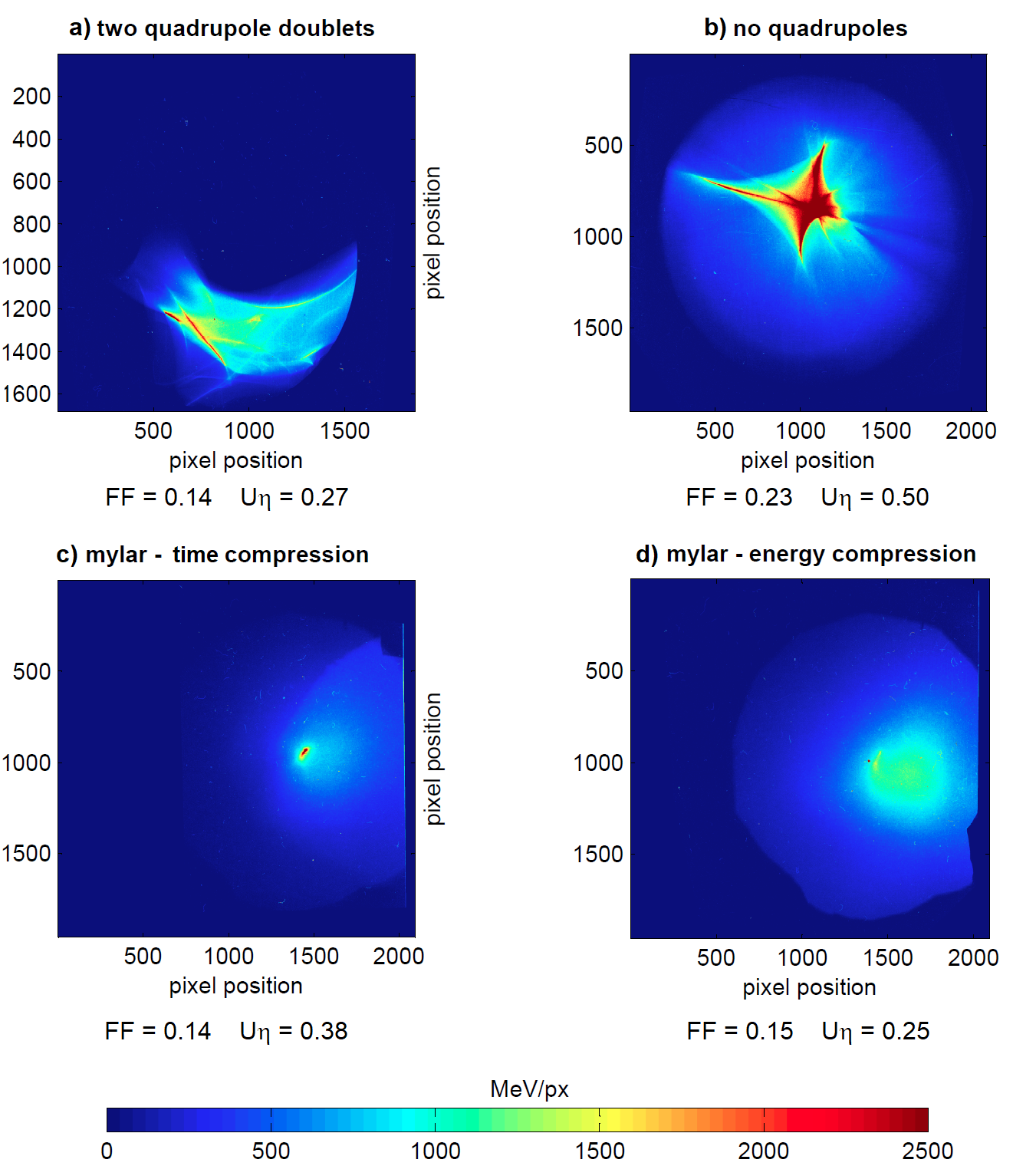}
\caption{Transverse beam profiles at 6~m distance from the target with the calculated fill factors (FF) and beam uniformities ($U_{\eta}$): a) beam with the installed quadrupole doublets behind the rf cavity, b) beam without the quadrupole doublets behind the cavity, c) time-compressed beam scattered with mylar foil, d) energy-compressed beam scattered with a mylar foil.}
\label{profiles}
\end{figure}
We measured the homogeneity with different beamline setups and settings in single shot measurements leading to the results shown in figure \ref{profiles}. In the old beamline design, two quadrupole doublets were installed after the rf cavity to support the particle transport leading to the fill factor FF~=~0.14 and the beam uniformity $U_{\eta}$~=~0.27. Based on beam shaping simulations with the TraceWin code from \textit{cea}\cite{Tracewin}, we removed the quadrupole doublets in our setup yielding a fill factor FF~=~0.23 and beam uniformity $U_{\eta}$~=~0.50. This profile covers a larger area leading to an improved fill factor. We observed a star-shaped, very intense and high-energetic spot leading to a worse beam uniformity. In the next step, we used a 1.25~$\mu$m thin mylar foil because of its neglegible energy-loss, for beam straggling to reduce the filametation. We measured the beam homogeneity in the energy-compressed (FF~=~0.15 and $U_{\eta}$~=~0.25) and time-compressed setting (FF~=~0.14 and $U_{\eta}$~=~0.38). Using the straggling foil, we significantly improved the beam homogeneity. In the case of energy-compression setting, the beam uniformity has the lowest value compared to the other setups. Nevertheless, the star-shaped spot became circular-shaped and still remained leading to a low fill factor.
In future, the collaboration will continue its work on improving the beam transport and the beam homogeneity.

\section{Proton imaging of a cold solid object}
One particular interest is in the imaging of macroscopic objects to study their properties \cite{mroth}. The dedicated LIGHT beamline offers high particle numbers, energy- or time-compressed proton bunches and a collimated beam which are excellent properties for imaging. As the first imaging sample we placed a solid target at the end of the beamline and used an energy-compressed proton beam with the mylar straggling foil (beam profile shown in figure 2d). \ \ \ \ \ \

\subsection{The imaging object}
\begin{figure}[htp!]
\includegraphics[scale=1]{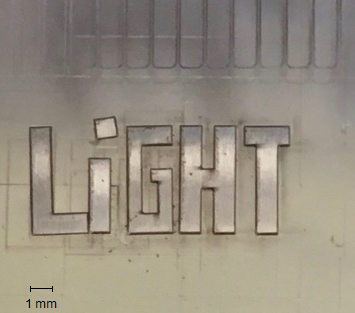}
\caption{Front view of the solid nickel imaging target with cut out letters.}
\label{logo}
\end{figure}

The imaging target consisted of six nickel foils in a stack. In each foil a certain number of letters was lasercut out of the layer resulting in different thicknesses which are listed in table \ref{tab}. Figure \ref{logo} shows a photograph of the designed target. Protons have a characteristic Bragg peak behaviour and deposit mainly their energy in the RCF foil, in which they are stopped. Due to the different thickness dependent on the local pixel position, the proton beam deposits energy within the imaging target leading to different energy distributions behind the target. The proton beam propagation through the imaging target was modeled with the MC simulation code \textit{SRIM} \cite{srim} which calculates the final spatial and energy distribution of ions passing through materials with a defined thickness. Based on \textit{SRIM}, the transmitted proton energy behind different letters was modeled, also listed in table \ref{tab}.
   
\begin{tiny}
\begin{table}[htp!]
\caption{Nickel foil thickness of different letters and transmitted proton energy of 8.5 MeV proton beam are calculated with \textit{SRIM}.}
\begin{tabular}{|c|c|c|}
\hline 
 & thickness in $\mu$m & transmitted energy in MeV \\ 
\hline 
i-dot & 0 & 8.5 \\   %8.0 
L & 20 &  7.9  \\  %7.4
I & 40 &  7.2 \\   %6.7 
G & 60 &  6.6  \\  %6.1
H & 80 &  5.8 \\   %5.2 
T & 100 & 5.1  \\   %4.3 
outside & 120 & 4.2  \\ % 3.4 \\ 
\hline 
\end{tabular} 
\label{tab}
\end{table}
\end{tiny}

\subsection{Proton imaging results}
\begin{figure}[htp!]
\includegraphics[scale=0.2]{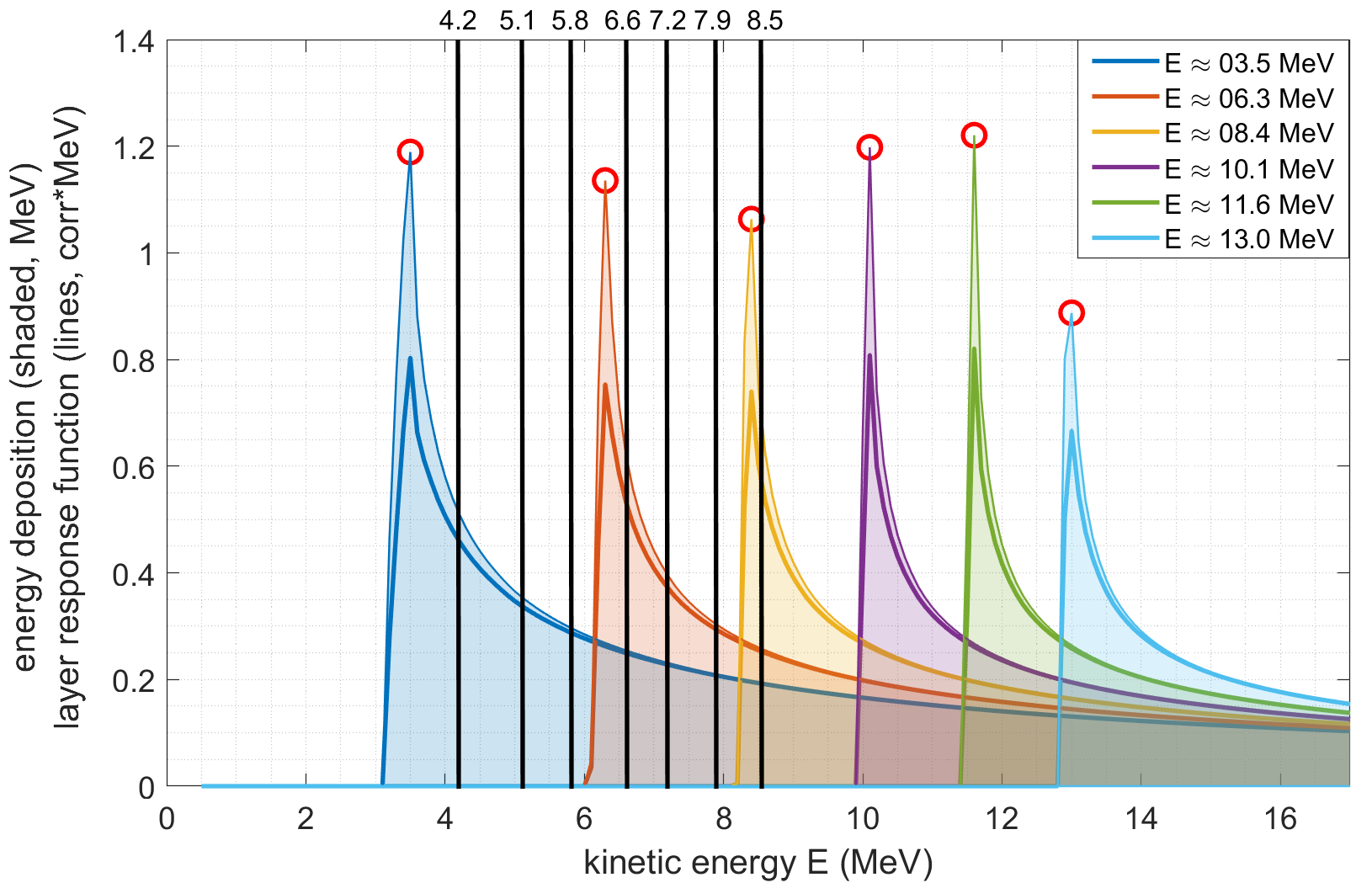}
\caption{Energy deposition and RCF response of protons in active layer: For each RCF layer the energy deposition in the active layer and the energy, at which the protons are stopped at their Bragg peak, are calculated. The expected \textit{SRIM} calculated transmitted energies are marked by black lines.}
\label{rcf}
\end{figure}

\begin{figure}
\includegraphics[scale=0.32]{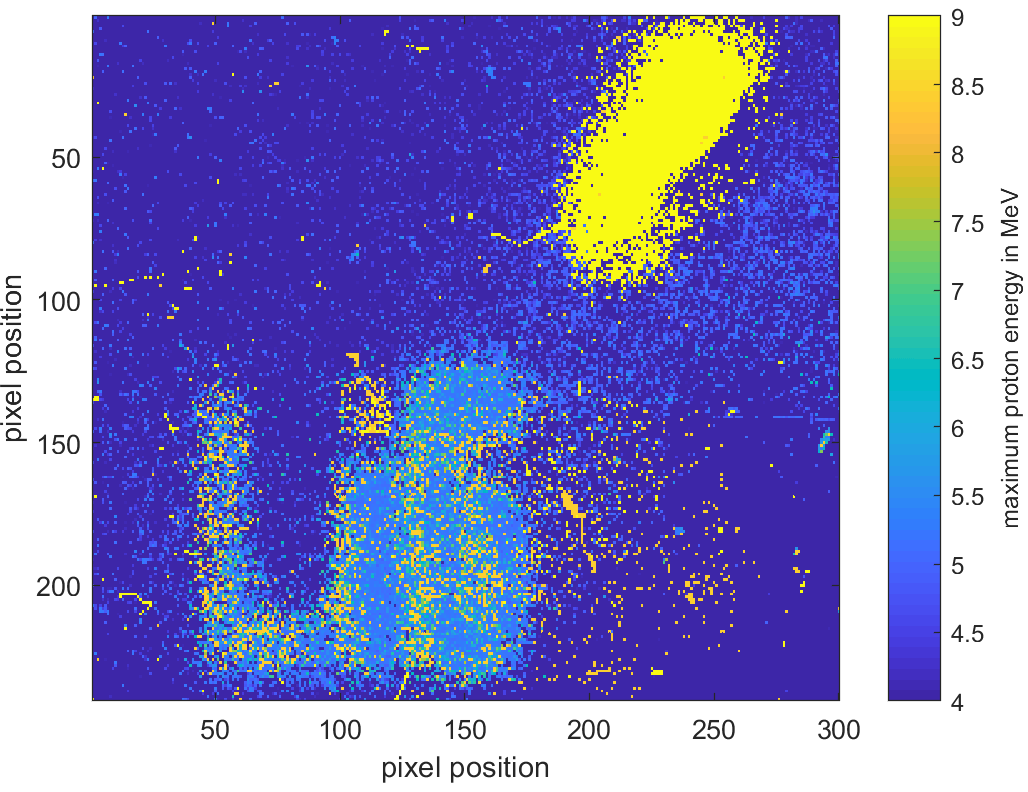}
\caption{Colour coding of the imaging results: The colour indicates the energy of the transmitted proton beam at each pixel: The unaffected beam passes the i-dot with an energy of 8.5 MeV. Behind the first three letters, the transmitted beam energy is in the range of 6.3 to 8.3 MeV. The last two letters and the surroundings are stopped in the first RCF layer. }
\label{code}
\end{figure}

The imaging object was positioned below the inhomogeneous spot (see figure \ref{profiles}) so it was hit by the almost homogeneous part of the beam. As detector a stack consisting of six spatially resolving RCF films (type:~EBT3) was used and positioned 6 mm behind the imaging object. Each EBT3 film consists of a 28 $\mu$m active layer sandwiched between 125 $\mu$m matte-surface polyester foils leading to an energy deposition shown in figure \ref{rcf}. In this figure, for each RCF layer the energy is shown, at which the protons are stopped within their Bragg peak. The expected transmission energies behind the imaging target are also indicated.  \\       
 The analysing code converted the deposited dose in the RCF films into an energy-dependent particle distribution for each pixel based on \textit{SRIM} calculated energy loss values. We determined the energy of the maximum for each fitted distribution and colour-coded this pixel indicating the transmitted beam energy. The colour-coding is shown in figure \ref{code}. The inhomogeneity, described in the previous section, is recognized above the letters. Although at this position the imaging target has a thickness of 120 $\mu$m, the transmitted proton beam has a higher energy at this position than at the reference position. Because of the straggling effects through the six nickel foils, the spot widened. The source of this observation will be investigated in the future. As reference for the unaffected beam, the protons passed the i-dot possessing a transmission energy of 8.5~MeV and are stopped as expected in the third RCF layer. The transmitted protons behind the first three letters, in the range of 20 to 60 $\mu$m nickel thickness, are stopped within the second RCF film so that their energies were in the range between 6.3 and 8.3~MeV. The calculated average transmitted energy was 6.5~MeV. Taking the characteristic proton behaviour (see figure \ref{rcf}) into account, behind the letter \glqq G\grqq \ the highest amount of energy was deposited, while behind the letters \glqq L\grqq \ and \glqq i\grqq \ the protons deposit less energy. The transmitted protons behind the letters \glqq H\grqq \ and \glqq T\grqq \ were stopped as well as the surrounding area in the first RCF layer having an average energy of 3.5~MeV. Because of the insufficient resolution between the different thicknesses, a more precise analysis was not possible. As the layers were positioned above each other with an accurary of 0.5 mm, the edges show different energies. The imaging results are agreeable with the expected energy loss values in table \ref{tab} and the transmitted proton are stopped within the expected RCF layers (see figure \ref{rcf}). In this first test, we showed that our beam is applicable for imaging techniques.

\section{Conclusion and outlook}
We successfully improved the proton beam homogeneity of the laser-driven LIGHT beamline through changing the beamline setting and using a mylar straggling foil. Nevertheless, a circular-shaped, high-energetic inhomogeneity remained in the transverse profiles. In future, we will work on an improved beam transport to increase the beam homogeneity. \\
In the second part, we placed a solid object for proton imaging at the end of the beamline and performed first imaging application studies. The proton imaging results are agreeable with the expected energy loss values of the transmitted beam, but a more precise analysis was limited by the RCF spectral resolution. In the next experimental campaign, the proton detector resolution will be improved by using a different type of RCF films (type: EBT3 with only one polyester substrate layer) which have a 28 $\mu$m active layer followed by a 125 $\mu$m matte-surface polyester foil. The missing substrate layer enables an increasement in resolution as more active layers will be in the desired energy range. This will enable a more precise analysis and we will be able to determine the density of the solid target.

% If you have acknowledgments, this puts in the proper section head.
\begin{acknowledgments}
The authors are thankful for the support by the PHELIX laser team, the HF group as well as the detector laboratory at GSI. This work is supported by HGS-HIRe.
\end{acknowledgments}

% Create the reference section using BibTeX:
\bibliography{bib_diana}
  \bibliographystyle{unsrt}

\end{document}